\documentclass[a4paper,10pt,twoside]{cpc-hepnp}

\usepackage{multicol}
\usepackage{graphicx}
\usepackage{booktabs}
\usepackage{amssymb,bm,mathrsfs,bbm,amscd}
\usepackage[tbtags]{amsmath}
\usepackage{lastpage}
\usepackage{CJK}

\begin{document}
\begin{CJK*}{GBK}{song}

\fancyhead[c]{\small Chinese Physics C~~~Vol. XX, No. X (201X)
XXXXXX} \fancyfoot[C]{\small 010201-\thepage}

\footnotetext[0]{Received 14 March 2009}

\title{Topological index associated with transverse axial vector and vector anomalies in QED\thanks{Supported by National Natural Science
Foundation of China (No.10775059) }}

\author{%
      SUN Yi-Qian$^{1;1)}$\email{windfall@gmail.com}%
\quad LU Pin$^{1;2)}$\email{pin@jlu.edu.cn}%
\quad BAO Ai-Dong$^{1,2,3;3)}$\email{baoad433@nenu.edu.cn}%
}
\maketitle

\address{%
$^1$ Center for Theoretical Physics, Department of Physics, Jilin University Chang chun 130023,china \\
$^2$ Department of Physics, Northeast Normal University, Chang chun 130024, China\\
$^3$ Department of Physics, Kashi Normal University, Xin jiang 844000, China\\}

\begin{abstract}
It is shown that a novel anomaly associated with transverse Ward-Takahashi identity exists for pseudo-tensor current in QED, and the anomaly gives rise to a topological index of Dirac operator in terms of Atiyah-Singer index theorem.
\end{abstract}



\footnotetext[0]{\hspace*{-3mm}\raisebox{0.3ex}{$\scriptstyle\copyright$}2013
Chinese Physical Society and the Institute of High Energy Physics
of the Chinese Academy of Sciences and the Institute
of Modern Physics of the Chinese Academy of Sciences and IOP Publishing Ltd}%

\begin{multicols}{2}

\section{Introduction}

Some time ago Takahashi made the argument for the plausible existence of transverse Ward Takahashi(WT) relation in canonical field theory, which has the potential to restrict the transverse vertex function from gauge symmetry alone[1]. Subsequently these transverse WT relations for the fermion-boson vertex in coordinate space (or in momentum space) are cast in four-dimensional Abelian gauge theory by computing the curl of the time ordered products of three-point Green functions [2,3]. In addition, the proposed transverse WT relation holds at one-loop order level in four dimensions gauge theory[4]. Up to the effect of quantum anomaly, the possible anomaly for the transverse Ward-Takahashi relations in four dimensional gauge theories is studies by He using the point-splitting method[5]. Recently, the anomaly issue reexamined by means of perturbative method. The conclusion is that there are no transverse anomalies for both the axial vector and vector current [6]. Also the path-integral derivation of the transverse WT relation for the vector vertex is presented due to a set of infinitesimal transformation of field variable in QED in Refs.[7,8], wherein topological property of the axial-vector current has been illustrated in terms of Atiyah-Singer index theorem. Based on the validity of Fujikawa's analysis, we reevaluate in detail the transverse anomaly of the transverse WT identity for the axial-vector and vector vertex in the QED, we find that a careful application of Fujikawa's approach leads to a transverse quantum anomaly for the axial vector current.[9]. In this paper, we point out that the mathematical explanation of such a anomaly is directly related to Atiyah-Singer index theorem. The topological index for Dirac operator corresponding to the anomaly will be expressed.

\section{Calculation of anomaly factor in Ward-Takahashi identity}
From the point of view of path-integral formulation, we proposed a infinitesimal transverse transformation of field variables to derive the WT identities[7][9]. Let us consider a set of infinitesimal local transformation in the QED

\begin{eqnarray}
\label{eq2}
\psi^{'}(x)&=&e^{-\frac{1}{4}\theta(x)\omega_{\mu\nu}\sigma^{\mu\nu}}\psi(x)\nonumber\\
\overline{\psi}^{'}(x)&=&\overline{\psi}(x)\gamma^{0}e^{-\frac{1}{4}\theta(x)\omega_{\mu\nu}\sigma^{\mu\nu\dagger}}\gamma^{0}\nonumber\\
B^{'}_{\mu}(x)&=&B_{\mu}(x)+\omega_{\mu\nu}\partial^{\nu}\theta(x).
\end{eqnarray}

where $\omega_{\mu\nu}$ stands for the antisymmetry tensor,$\psi(x)$ and $B_{\mu}(x)$ are the fermion and gauge fields,respectively. \\
In principle, the variation of the generating functional itself under the transformation of field variables Eq.$(1)$ can lead to Ward-Takahashi type's identities  .The change of the function integral due to the transformation (choosing $\delta B_{\mu}(x)=0 $ for simplicity) gives the relation  in momentum space in QED
\begin{eqnarray}
&-iq^{\mu}\Gamma^{\nu}_{V}(p_{1},p_{2})+iq^{\nu}\Gamma^{\mu}_{V}(p_{1},p_{2})+(p_{1\lambda}+p_{2\lambda})\nonumber\\
&\times\varepsilon^{\lambda\mu\nu\rho}q_{\lambda}\Gamma_{A\rho}(p_{1},p_{2})+
+S^{-1}(p_{2})\sigma^{\mu\nu}+S^{-1}(p_{1})\sigma^{\mu\nu}\nonumber\\
&-\int\frac{d^{4}k}{(2\pi)^{4}}2k_{\lambda}\epsilon^{\lambda\mu\nu\rho}\Gamma_{A\rho}(p_{1},p_{2};k)=0
.
\end{eqnarray}
This WT relation for the vector current has been listed in Ref.[9].
The integral term in Eq.(2) may be called the integral-term involving the vertex function $\Gamma_{A\rho}(p_{1},p_{2};k)$ with the internal momentum $k$ of the gauge boson appearing in the Wilson line$[10]$. The Fourier transformation for vertex function $\Gamma_{A\rho}(p_{1},p_{2};k)$ is defined as
\end{multicols}
\begin{eqnarray}
\int~d^{4}xd^{4}x_{1}d^{4}x_{2}e^{i(p_{1}x-p_{2}x_{2}-qx)}\langle0|T(\overline{\psi}(x)\varepsilon^{\lambda\mu\nu\rho}\gamma_{\rho}\gamma^{5}\psi(x)\overline{\psi}(x_{1})\psi(x_{2})U_{P}(x^{'},x))|0\rangle\nonumber\\
=(2\pi)^{4}\delta^{4}(p_{1}-p_{2}-q)iS_{F}(p_{1})\Gamma_{A\rho}(p_{1},p_{2},k)iS_{F}(p_{2}).
\end{eqnarray}
\begin{multicols}{2}
where $ q=(p_{1}-k)-(p_{2}-k)$.
Obviously, the full vector function and the full axial -vector function are coupled with each other.As shown is Ref[3], the apparent feature of this transverse identity Eq.(2) is that the vertex function $\Gamma^{\mu}_{V}$ (fermion's three point function)  has the transverse component of itself .
\par\indent Completely analogous to the calculations above, let us consider the other transverse transformation
\begin{eqnarray}
\psi^{'}(x)&=&e^{-\frac{1}{4}\theta(x)\omega_{\mu\nu}\sigma^{\mu\nu}\gamma^{5}}\psi(x)\nonumber\\
\overline{\psi}^{'}(x)&=&\overline{\psi}(x)\gamma^{0}e^{-\frac{1}{4}\theta(x)\omega_{\mu\nu}\gamma^{5\dagger}}\sigma^{\mu\nu\dagger}\gamma^{0}
\end{eqnarray}
we obtain the identity for the axial -vector current in momentum space as
\end{multicols}
\begin{eqnarray}
&-iq^{\mu}\Gamma^{\nu}_{A}(p_{1},p_{2})+iq^{\nu}\Gamma^{\mu}_{A}(p_{1},p_{2})+(p_{1\lambda},p_{2\lambda})\varepsilon^{\lambda\mu\nu\rho}q_{\lambda}\Gamma_{V\rho}(p_{1}+p_{2})+\nonumber\\
&+S^{-1}(p_{2})\sigma^{\mu\nu}\gamma^{5}-S^{-1}(p_{1})\sigma^{\mu\nu}\gamma^{5}-\int\frac{d^{4}k}{(2\pi)^{4}}2k_{\lambda}\epsilon^{\lambda\mu\nu\rho}\Gamma_{V\rho}(p_{1},p_{2};k)=0.
\end{eqnarray}
\begin{multicols}{2}
According to Fujikawa's interpretation, it is argued that the appearance of the quantum anomaly in WT identity is a symptom of the impossibility of defining a suitably invariant functional integral measure due to the relevant transformations on fermionic field variables.The regularization procedure for the variations of the integral measure can provide access to a wider class of such anomaly objects [11][12][13][14].To see how the change of the measure corresponding to the transverse transformation Eq.(1)and Eq.(4) gives rise to a possible anomaly factor, let us consider an Abelian gauge field to show our argument. The Lagrangian density for massive QED, which is of the form
\end{multicols}
\begin{eqnarray}
L_{eff}=\overline{\psi}(x)\gamma^{\mu}(\partial_{\mu}-ieB_{\mu}(x))\psi(x)-\overline{\psi}(x)m\psi(x)-
   \frac{1}{4}F^{\mu\nu}F_{\mu\nu}-\frac{1}{2\xi}(\partial^{\mu}B_{\mu})^{2}.
\end{eqnarray}
\begin{multicols}{2}
where $e$ and $m$ denote, respectively, the charge and mass of the electron. In this case, the gauge field is just the photon field $B_{\mu}(x)$.
 Thus jacobian $J^{[\alpha\beta]}(x)$ of integral measure due to the transformations Eq.$(1)$ is evaluated  below
\begin{eqnarray}
J^{[\alpha\beta]}(x)=e^{\int~d^{4}x A^{[\alpha\beta]}(x)\omega_{[\alpha\beta]}\theta(x)}e^{\int~d^{4}(x)\bar{A}^{[\alpha\beta]}(x)\omega_{[\alpha\beta]}\theta(x)}.
\end{eqnarray}
This is what we set out to calculate.
  Due to the transverse transformation(2.1), the  anomaly functions  can be written as the limit of a manifestly convergent integral
\end{multicols}
\begin{eqnarray}
A^{[\alpha\beta]}(x)&=&\underset{M\to\infty}{lim}\int\frac{d^{4}k}{(2\pi)^{4}}e^{-ikx}(\frac{-\sigma^{\alpha\beta}}{4})f\big((\frac{i\gamma^{\mu}D_{\mu}}{M})^{2}\big)e^{ikx}\nonumber\\
\overline{A}^{[\alpha\beta]}(x)&=&\underset{M\to\infty}{lim}\int\frac{d^{4}k}{(2\pi)^{4}}e^{-ikx}\gamma^{0}(\frac{-\sigma^{\alpha\beta}}{4})^{\dagger}\gamma^{0}f\big((\frac{i\gamma^{\mu}D_{\mu}}{M})^{2}\big)e^{ikx}.
\end{eqnarray}
\begin{multicols}{2}
where $D_{\mu}=\partial_{\mu}-ieB_{\mu}(x)$ is the covariant derivative.
In addition, the transformation of the field $B_{\mu}(x)$   is a translation, so that its Jacobian is trivial.
The anomaly function $ A^{[\alpha\beta]}(x)$ requires regulation,which is achieved by inserting a regulator
\begin{eqnarray}
f\big(\frac{-(\gamma^{\mu}D_{\mu})^{2}}{M^{2}}\big)&=&\big(e^{-\frac{(\gamma^{\mu}D_{\mu})^{2}}{M^{2}}}\big)^{2}
\end{eqnarray}
The expression of the transverse vector anomaly function $A^{[\alpha\beta]}(x)$ can be put in the regulating form
\end{multicols}
\begin{eqnarray}
 A^{[\alpha\beta]}(x)&=&\underset{M\to\infty}{lim}\int\frac{d^{4}k}{(2\pi)^{4}}e^{-ikx}(\frac{-1}{4}\sigma^{\alpha\beta})f\big((\frac{i\gamma^{\mu}D_{\mu}}{M})^{2}\big)e^{ikx}\nonumber\\
 &=&\underset{M\to\infty}{lim}\int\frac{d^{4}k}{(2\pi)^{4}}e^{-ikx}\big[(\frac{-1}{4}\sigma^{\alpha\beta})(\sum_{n}\frac{1}{n!}f^{(n)}(-\frac{D^{2}}{M^{2}})(\frac{i}{4M^2}[\gamma^{\mu},\gamma^{\nu}]F_{\mu\nu})^{n})\nonumber\\
&&\times\big[\sum_{m}\frac{1}{m!}f^{(m)}(-\frac{D^{2}}{M^{2}})(\frac{i}{4M^2}[\gamma^{\rho},\gamma^{\sigma}]F_{\rho\sigma})^{m}]e^{ikx}\nonumber\\
 &=&\underset{M\to\infty}{lim}\int\frac{d^{4}k}{(2\pi)^{4}}Tr\big[(\frac{-1}{4}\sigma^{\alpha\beta})(f^{(')}(-\frac{D^{2}}{M^{2}})(\frac{i}{4M^2}[\gamma_{\mu},\gamma_{\nu}]F_{\mu\nu})^{1})(f^{(')}(-\frac{D^{2}}{M^{2}})(\frac{i}{4M^2}[\gamma_{\rho},\gamma_{\sigma}]F_{\rho\sigma})^{1})\nonumber\\
&=&\frac{1}{128\pi^{2}}\big(g^{\alpha\mu}g^{\beta\rho}g^{\nu\sigma}-g^{\alpha\mu}g^{\beta\sigma}g^{\nu\rho}+g^{\alpha\nu}g^{\beta\sigma}g^{\mu\rho}-g^{\alpha\nu}g^{\beta\rho}g^{\mu\sigma}+\nonumber\\
&&+g^{\beta\mu}g^{\alpha\sigma}g^{\nu\rho}-g^{\beta\mu}g^{\alpha\rho}g^{\nu\sigma}+g^{\beta\nu}g^{\alpha\rho}g^{\nu\sigma}-g^{\beta\nu}g^{\alpha\sigma}g^{\mu\rho}\big)Tr(F_{\mu\nu}F_{\rho\sigma}).
\end{eqnarray}
\begin{multicols}{2}
In terms of the symmetry of metric and antisymmetry of 4-dimensional field strength tensor, we expand the anomaly function and find that it equals zero. Thus the Jacobian Eq.$(7)$ becomes
\begin{eqnarray}
J^{[\alpha\beta]}(x)&=&e^{2\int~d^{4}x A^{[\alpha\beta]}(x)\omega_{[\alpha\beta]}\theta(x)}\nonumber\\
&=&1.
\end{eqnarray}
\par
By the parallel procedure, for the case of the transformation Eq.(4), the transverse axial vector anomaly function  is given by
\end{multicols}
\begin{eqnarray}
A^{[\alpha\beta5]}(x)&=&\underset{M\to\infty}{lim}\int\frac{d^{4}k}{(2\pi)^{4}}e^{-ikx}(\frac{-1}{4}\sigma^{\alpha\beta}\gamma^{5})f\big((\frac{i\gamma^{\mu}D_{\mu}}{M})^{2}\big)e^{ikx}\nonumber\\
&=&\underset{M\to\infty}{lim}\int\frac{d^{4}k}{(2\pi)^{4}}e^{-ikx}\big[(\frac{-1}{4}\sigma^{\alpha\beta}\gamma^{5})(\sum_{n}\frac{1}{n!}f^{(n)}(-\frac{D^{2}}{M^{2}})(\frac{i}{4M^2}[\gamma_{\mu},\gamma_{\nu}]F_{\mu\nu})^{n})\nonumber\\
&&\times\big[\sum_{m}\frac{1}{m!}f^{(m)}(-\frac{D^{2}}{M^{2}})(\frac{i}{4M^2}[\gamma^{\rho},\gamma^{\sigma}]F_{\rho\sigma})^{m}]e^{ikx}\nonumber\\
&=&\underset{M\to\infty}{lim}\int\frac{d^{4}k}{(2\pi)^{4}}Tr\big[(\frac{-1}{4}\sigma^{\alpha\beta}\gamma^{5})(f^{(')}(-\frac{D^{2}}{M^{2}})(\frac{i}{4M^2}[\gamma^{\mu},\gamma^{\nu}]F_{\mu\nu})^{1})(f^{(')}(-\frac{D^{2}}{M^{2}})(\frac{i}{4M^2}[\gamma^{\rho},\gamma^{\sigma}]F_{\rho\sigma})^{1})\nonumber\\
&=&\frac{i}{128\pi^{2}}\big(-g^{\alpha\mu}\epsilon^{\beta\nu\rho\sigma}+g^{\alpha\nu}\epsilon^{\beta\mu\rho\sigma}+g^{\beta\mu}\epsilon^{\alpha\nu\rho\sigma}-g^{\beta\nu}\epsilon^{\alpha\mu\rho\sigma}+\nonumber\\
&&+g^{\nu\rho}\epsilon^{\alpha\beta\mu\sigma}-g^{\mu\rho}\epsilon^{\alpha\beta\nu\sigma}+g^{\mu\sigma}\epsilon^{\alpha\beta\nu\rho}-g^{\nu\sigma}\epsilon^{\alpha\beta\mu\rho}\big)Tr(F_{\mu\nu}F_{\rho\sigma})\nonumber\\
&=&-\bar{A}^{[\alpha\beta5]}(x).
\end{eqnarray}
\begin{multicols}{2}
The corresponding Jacobian is non-trivial in the form\\
\begin{eqnarray}
J^{[\alpha\beta5]}(x)=e^{2\int~d^{4}x A^{[\alpha\beta]}(x)\omega_{[\alpha\beta]}\theta(x)}.
\end{eqnarray}

Obviously the  Eq.(11) and Eq.(13) is perfectly consistent with result of derivation of transverse vector and axial vector anomalies in four-dimensional $U(1)$ gauge theory using perturbative methods[5]. \\

\section{Topological index for the anomaly}
From the topological viewpoint, the topological property of the quantum anomaly Eq.(13) is addressed by Atiyah-Singer index theorem for Dirac operator in Eq.(6) in a gauge background [15]. Since $D$ is an elliptic operator, the Atiyah-Singer local index formula for the topological index $Ind_{t}D$ of $D$ in terms of the chern character of $ChE$ and the $\widehat{A}$-genus of $E^4$ gives
\begin{eqnarray}
Ind_{a}D &=&Ind_{t}D \\
&=&\int \widehat{A}\left( M\right) \cdot Ch\left( E\right).
\end{eqnarray}

Here the more natural definition of$\widehat{A}$ -genus form with the Riemannian curvature $R$ is

\begin{eqnarray}
\widehat{A}\left(   M\right) =det^{\frac{1}{2}}\frac{\frac{R}{2}}{\sinh \left(
\frac{R}{2}\right) }.
\end{eqnarray}
And the chern character form is

\begin{eqnarray}
Ch\left( E\right) =Str\left( e^{-\triangledown ^{2}}\right).
\end{eqnarray}

Substituting this into Eq.(14) we have

\begin{eqnarray}
Ind_{t}D=-\frac{4}{32\pi ^{2}}\int d^{4}xTr\partial _{\mu }\left(
\varepsilon ^{\mu \nu \rho \sigma }B_{\nu }\partial _{\rho }B_{\rho }\right).
\end{eqnarray}

Obviously the quantity on the right of the Eq.(18) is known as chen-pontrjagin term.The Eq.(18) can be expressed by a complete of eigenstates  $\phi_{n}$
of Dirac operater $D$
\begin{eqnarray}
Ind_{t}(D)&=&\underset{M\to\infty}{lim}\int d^{4}x\langle x|\Gamma ^{[5]} f(-\frac{D^{2}}{M^{2}})|x\rangle\nonumber\\
&=&\int d^{4}xA^{[5]}(x).
\end{eqnarray}
Here,the result in Eq.(19) is no other than the topological index for the corresponding anomaly function $A^{[5]}(x)$  of axial-vector current[8].
Furthermore we generalize the above approach to the pseudo-tensor current.
According to the local Atiyah-Singer index theorem, the existence of an asymptotic expansion for $\langle x|\Gamma^{[\alpha\beta 5]}  f(-D^2/M^2)|x \rangle$ at $M\to\infty $ is presented by the index $ Ind(\Gamma^{[\alpha \beta 5]} , D) $ of Dirac operator $D$ on the manifold $M$ , Which is calculated below
\begin{eqnarray}
Ind(\Gamma ^{[\alpha\beta5]}, D)&=&\underset{M\to\infty}{lim}\int d^{4}x\langle x|\Gamma ^{[\alpha\beta5]} f(-\frac{D^{2}}{M^{2}})|x\rangle\nonumber\\
&=&\int d^{4}xA^{[\alpha\beta5]}(x).
\end{eqnarray}

Here the kernel function $A^{[\alpha\beta5]} (x)$ in Eq.(20) is no other than the anomaly function Eq.(12) for Jacobian of functional measure Eq.(8).
This is what we do. It is shown that the index theorem for the operator $\Gamma^{[\alpha\beta5]}  e^{(-D^2/M^2 )}$ associated with the square of Dirac operator D is a statement about the relationship between the kernel $A^{[\alpha\beta5]}(x)$ and the topological character in gauge theory.
Fortunately, the asymptotic expansion of the kernel $A^{[\alpha\beta]}(x)$ of the operator  $\Gamma^{[\alpha\beta5]} e^{(-D^2/M^2 )}$ on the right of Eq.(5) has evaluated in the Eq.(12) by Fujikawa's method for these fermion pseudo-currents.

\section{concluding Remarks}
As already described, the topological index of Dirac operator   reflecting the property of fermion current coupling gauge field is indeed understand on Atiyah-Singer index theorem in quantum field theory as a consequence of the fact that the Jacobian of integration measure possesses anomaly terms, which is associated with Ward-Takahashi identity for pseudo- tensor current in QED. \\

\section{References}

\end{multicols}

\vspace{10mm}
\begin{multicols}{2}

\end{multicols}

\clearpage

\end{CJK*}

\begin{thebibliography}{90}

\vspace{3mm}


\bibitem{lab1}Takahashi Y, Mancini F. In Quantum Field Theory, Amsterdam: Elsevier Scince Publisher,  1986, 19
\bibitem{lab2} KondoK-I,Maris.  Phys.Rev.D,1995, { \bf{52}:} 1212---1233
\bibitem{lab3} HE H X., Khanna F C and Takahashi Y.   Phys.Lett.B, 2000,{\bf{480}:} 222---228
\bibitem{lab4}Pennnington M R , Williams R.   J.Phys.G:Nucl.Part.Phys., 2006{\bf{32}:}2219---2233
\bibitem{lab5} He H X.   Phys.Lett.B, 2001, {\bf{507:}},351---355
\bibitem{lab6} SUN W S , ZONG H S , CHEN X S  et al.  Phys. Lett.B,2003, {\bf{569(2)}:},211---238
\bibitem{lab7}BAO A D , Wu S S.  Int.J.Theor.Phys., 2007,{\bf{46}:}12---27
\bibitem{lab8}BAO A D ,YAO H B  and WU S S. Chinese Phys.C,2009, {\bf{33(3)}:},177---182
\bibitem{lab9} BAO A D ,SUN Y Q,  and WAN D. Chin. Phys.Lett, 2012, {\bf{29}:},121101---121104
\bibitem{lab10}HE H X.   An Introduction to Nuclear Chromodynamics ,China, China University of Sci and Tech Press, 2009. 12
\bibitem{lab11} HE H X.  Int.J.Mod.Phys.A, 2007,{\bf{22:}},2119---2132
\bibitem{lab12} Fujikawa K.   Phys.ReV.Lett.  1997,{\bf{42}:},1195---1198
\bibitem{lab13}EinhornM B ,T Jones D R.  Phys.Rev.D, 1984 {\bf{29}:},331---333
\bibitem{lab14}Umezswa M  ,  Phys.Rev.D ,1989,{\bf{39}:},3672---3683
\bibitem{lab15}Nicole Verline, Ezra Getzle and Michele, Vergne.  Heat Kernels and Dirac operators,  Veilag Berlin Heidelberg: Springer 2004.18--169






\end{thebibliography}
\end{document}